\documentclass{aip-cp}

\usepackage[numbers]{natbib}
\usepackage{rotating}
\usepackage{graphicx}
\usepackage{tikz}
\usetikzlibrary{calc,shapes}

\newcommand{\tikzmark}[1]{\tikz[overlay,remember picture] \node (#1) {};}
\newcommand{\DrawBox}[1][red]{%
    \tikz[overlay,remember picture]{
    \draw[#1]
      ($(bl)+(-0.4em,1.8em)$) rectangle
      ($(br)+(0.4em,-1.2em)$);}
}
\newcommand{\MyBox}[2][black]{\tikzmark{bl}#2\tikzmark{br}\DrawBox[#1]}

\begin{document}

\title{Trade-off between sensitivity and selectivity in olfactory receptor neuron}

\author[aff1]{Alexander Vidybida\corref{cor1}
}
\eaddress[url]{http://vidybida.kiev.ua}
\affil[aff1]{Bogolyubov Institute for Theoretical Physics, Metrologichne str. 14-B, 03680 Kyiv, Ukraine}
\corresp[cor1]{Corresponding author: vidybida@bitp.kiev.ua}

\maketitle

\begin{abstract}
It was observed before that due to convergence in the olfactory system
 a possible amplification can be as large as the degree of convergence. 
 This is in the case when a single impulse
from the converging inputs is enough to trigger the secondary neuron. On the other hand,
if a number of impulses are required for triggering, a gain in discriminating ability
may be obtained along with decrease in sensitivity gained due to the convergence. 
We discuss this trade-off in terms
of concrete estimates using olfactory sensory neuron and the set of its receptor proteins
as an example of system with convergence.
\end{abstract}

\section{INTRODUCTION}
\begin{figure}[h]
  \centerline{\includegraphics[angle=-90,width=0.4\textwidth]{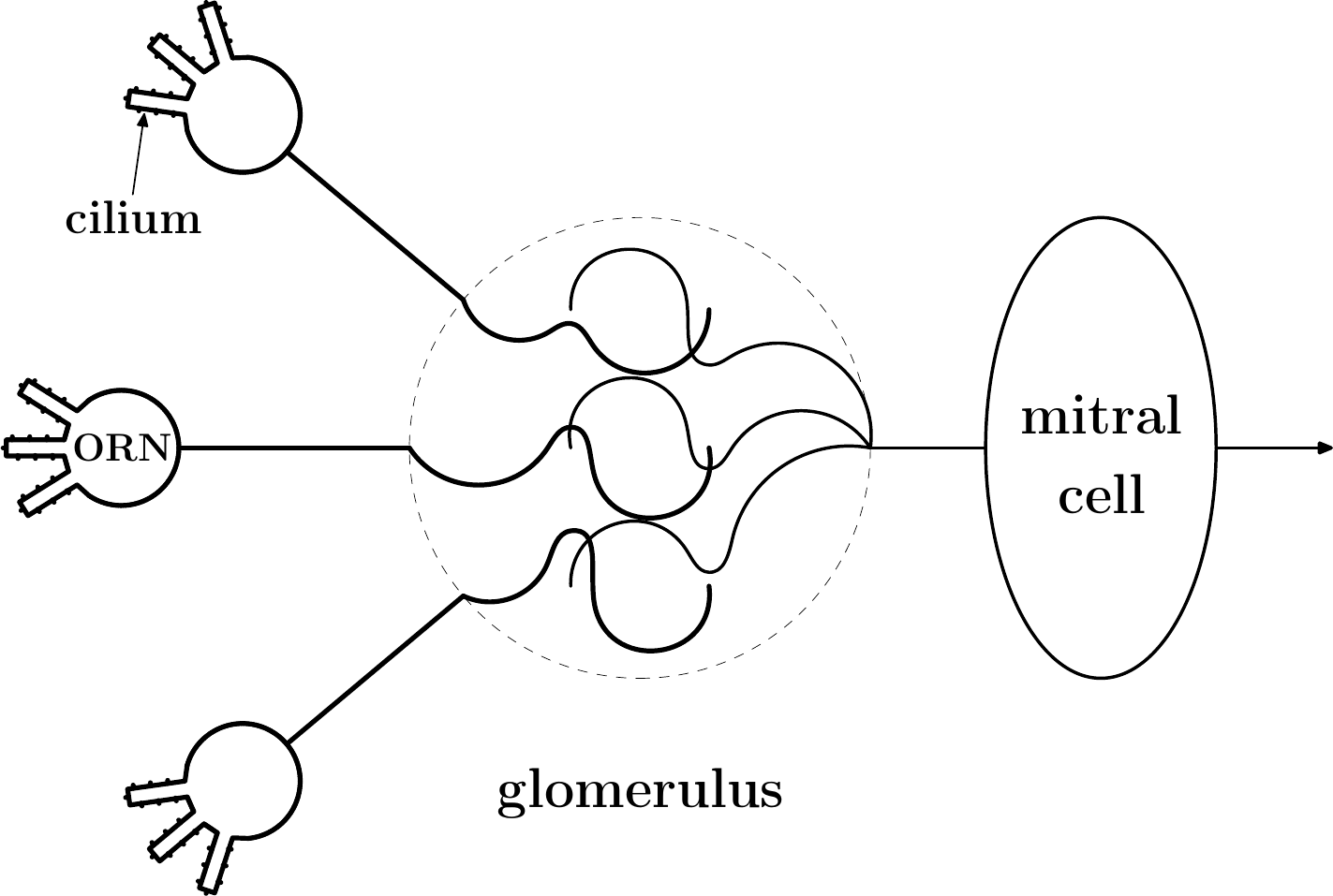}}
  \caption{\label{conver}Schematic example of convergent structure in olfactory system. Here as much as several thousands \citep{Buck2000} (the concrete number is species-dependent) of olfactory sensor
   neurons may converge  through a single glomerulus onto a single mitral cell.}
\end{figure}

In a neural system, a convergent organization of inter-neuronal connections is a typical pattern.
E.g., the degree of convergence observed for rat hippocampal pyramidal cells is about 12\,000,
 \cite{Andersen1990}. In the olfactory system of mouse, see Fig. \ref{conver}, 
 about 5\,000 olfactory receptor neurons (ORN) send their output spikes to a single mitral cell,
  \cite{Ressler1994}. Our purpose is to figure out what role the convergence may have
  in forming sensitivity and selectivity in a sensory system. For this purpose we use olfactory
  system as an example.

\section{PROBABILISTIC DESCRIPTION}
Real neurons are noisy \cite{Rolls2010}. This has a consequence for a sensory system under 
weak stimulation. Namely, if a stimulus has low level, the response to it becomes probabilistic.
The dose-response dependence becomes probabilistic as well: the higher is the dose, the higher
is the response probability,  \cite{Drongelen1978,Drongelen1978a}. The sensitivity is considered
to be higher when the same stimulus evokes a response (output spike) with higher probability.

\section{SENSITIVITY GAIN DUE TO CONVERGENCE}
It was observed, \cite{Kaissling1970}, that threshold odor concentration, which is
able to evoke a response at the level of
whole animal can be 10-100 times lower than that at the level of ORN. This might be due to
convergence in the olfactory sensory pathways. 
At the stage of ORN-mitral cell communication, the explanation has been proposed
by W. van Drongelen {\em et. al.}, \citep{Drongelen1978,Drongelen1978a}. Namely, if under
a certain stimulation (odor concentration) an ORN output activity can be modeled as Poisson
stochastic process with intensity $\lambda$, then compound input of $N$ ORN outputs into a single
mitral cell can be modeled as Poisson process of intensity $N\lambda$.
The probability that a single ORN emits a spike during time interval $T$ 
is $P_{ORN}=1-e^{-\lambda T}\approx \lambda T$. The latter approximation is valid
for a short enough $T$. Expect that a single input spike is able to trigger
the secondary neuron. Then the probability that the secondary neuron emits a spike during $T$ 
is $P_{m}=1-e^{-N\lambda T}\approx N\lambda T$. In this case we have sensitivity gain
equal to $\frac{N\lambda T}{\lambda T}=N$. Actually, a real mitral cell requires more than a
single input impulse for triggering. Denote the required for triggering number of input
impulses as $N_0$. It can be proven that the 
mitral cell output rate $R_m$ in this case satisfies the following estimate:
$$
R_m\le \frac{N}{N_0}\lambda.
$$
The equality holds here if the mitral cell is modeled as perfect integrator. The corresponding
sensitivity gain $G_n$ is as follows:
\begin{equation}\label{Gn}
G_n=\frac{R_m}{\lambda}\le\frac{N}{N_0}.
\end{equation}

\section{SELECTIVITY GAIN INSIDE THE OLFACTORY RECEPTOR NEURON}

The olfactory system has the remarkable capacity to
discriminate among a wide range of odor molecules.
This begins with the ORN,
which performs the task of converting information 
contained in the odor molecules into information 
contained in membrane signals and neural space,
\cite{Shepherd1994}. Discriminating ability of ORN starts to build up
 already at the level of individual receptor proteins.
The primary act of odor perception happens when odor molecule 
physically contacts with a receptor protein integrated in the ORN's membrane.
During this contact, the molecule can be bound to the receptor with some probability
$p$. The probability $p$ depends on the chemical nature of the molecule,
which determines its affinity to a given receptor. It is namely due to this dependence
that an ORN is able to discriminate between different odors.

\subsection{Selectivity of Single Receptor Protein}
The discriminating ability of a receptor protein can be defined as follows.
Let during two separate experiments two different odors O and O' are presented
to a given receptor or a set of them at the same concentration. Denote as $p$
and $p'$ the probability to find a receptor bound with odor O and O'.
If $p\ne p'$ then we say that this receptor is able to discriminate between
O and O'. Otherwise, it cannot. The same is with the ORN expressing this receptor.
Suppose that $p>p'$. Then the discriminating
ability at the level of single receptor protein can be characterized numerically
as
\begin{equation}\label{mu}
 \mu = \log\frac{p}{p'}.
\end{equation}

\subsection{Selectivity of ORN}
\begin{figure}[h]
  \centerline{\includegraphics[width=0.3\textwidth]{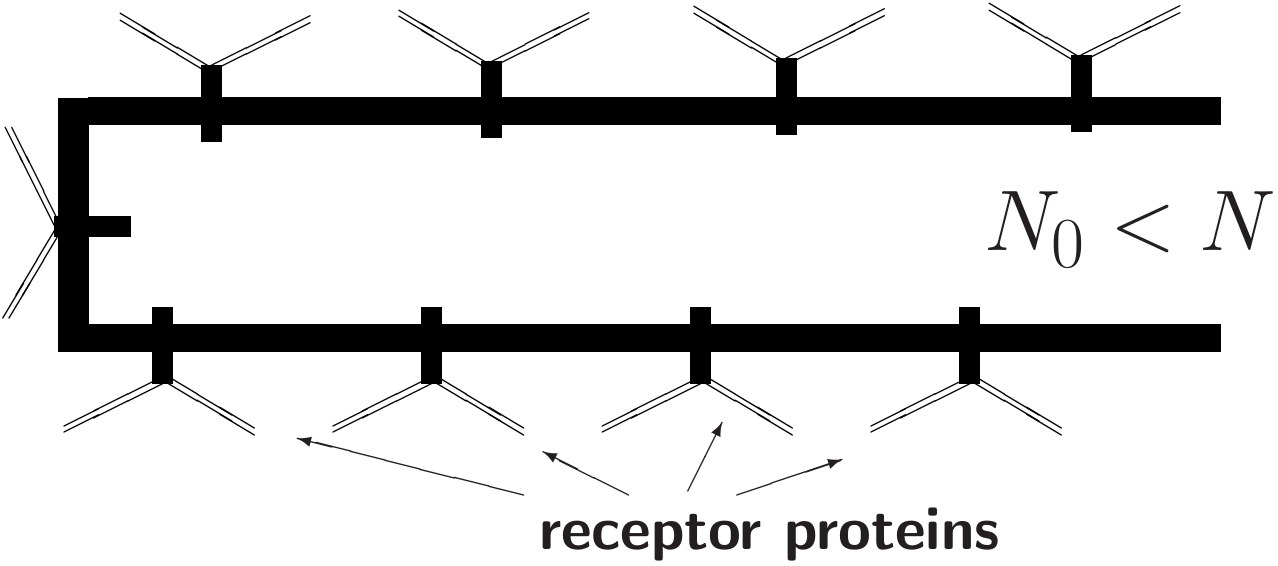}}
  \caption{\label{ORN}Simplified model of olfactory receptor neuron as single cilium
  with firing threshold. 
  $N$ --- is the total number of receptors, $N_0$ --- is the threshold number.
  The neuron starts firing if the number of bound receptors exceeds $N_0$.}
\end{figure}

In the ORN expressing concrete receptor protein, one may observe convergence 
of signals from individual receptor proteins onto the ORN's interior,
see Fig. \ref{ORN}, and
farther onto the axonal hillock where it is decided whether to fire or not.
The degree of convergence is characterized by the total number $N$
of receptors per neuron.
The firing threshold can be expressed as the minimal number $N_0$ of
receptors which must be bound with odor in order to ensure depolarization necessary
for triggering. Since odor binding-releasing is driven by the Brownian
motion, the firing threshold will be achieved irregularly and this will
result in irregular/random spiking of ORN.

In a simplified model of ORN, see Fig. \ref{ORN}, the ORN's discriminating
ability between the O and O' can be defined as follows.
Let the ORN during two separate experiments is exposed to O and O' at equal
concentrations. Denote $\overline{F}$ and $\overline{F'}$ the ORN's
mean firing rate if O and O' are presented. Then the ORN's ability $\delta$
to discriminate between O and O' can be defined as follows
\begin{equation}\label{delta}
\delta=\log\frac{\overline{F}}{\overline{F'}}.
\end{equation}

It was shown, \cite{Vidybida1991}, that in situation when the dose-response dependence is 
of threshold type and probabilistic, 
a considerable selectivity gain is possible. 
For the model ORN, the following mathematically rigorous estimate has been obtained,
\cite{Vidybida1999,Vidybida2000,Vidybida2003a}:
\begin{equation}\label{finest}
\MyBox{
\delta>\frac{p_0-p}{1-p}\,\,{N\,\mu},\quad p_0=\frac{N_0}{N}\,.
}
\end{equation}\medskip

\noindent
The numbers $N$ and $N_0$ can be high. E.g., $N=2\,500\,000$, $N_0=250$ for moth,
(J.-P. Rospars, private communication). For frog, $N=25\,000$, $N_0=35$, 
\cite{Bhandawat2010}.
Estimate (\ref{finest}) suggests that
ORN's selectivity can be much higher than that of its receptor proteins,
provided $p_0-p>0$. The latter is achieved if the odors O and O' are presented 
at sub-threshold concentration. Namely, the mean number of bound receptors is below
the threshold one. In this case, the firing threshold is achieved due to
random fluctuations, and this process appeared to be more selective than random
binding-releasing at a single receptor. Conclusion (\ref{finest}) has been checked
by means of direct numerical simulation of odor binding-releasing in a set of
receptors, \cite{Vidybida2008b}.

\section{TRADE-OFF BETWEEN SENSITIVITY AND SELECTIVITY}

Define the selectivity gain, $G_l$, as follows
\begin{equation}\label{Gl}
G_l = \delta/\mu.
\end{equation} 
Then from (\ref{Gn}) and (\ref{finest}) one obtains the following estimate
\begin{equation}\label{Gl1}
G_l>\frac{\frac{1}{G_n}-p}{1-p}N.
\end{equation}
The right-hand side of inequality (\ref{Gl1}) is the decreasing function of $G_n$.
Therefore, the better is the sensitivity gain the less optimistic are estimates
(\ref{finest}), (\ref{Gl1}) for selectivity. 
Consider situation when all $N$ 
receptors should be bound to ensure triggering ($N_0=N$). In this case $G_n=1$ 
(no sensitivity gain) and we have from (\ref{Gl1})
$$
G_l>N,
$$
which is quite optimistic for selectivity.

The estimate (\ref{Gl1}) depends on the probability $p$ that a receptor is bound
with odor O. This probability can be calculated based on association-dissociation
rate constants $k_+$, $k_-$ between the odor and receptor, and odor's concentration, $c$:
$$p=\frac{c\,k_+}{c\,k_+\,+k_-}.$$
If concentration $c$ is very low, then $p$ is very low as well and
(\ref{Gl1}) turns into the following:
$$
G_l\,G_n>N,
$$
which again demonstrates the trade-off between selectivity and sensitivity.
Finally, if concentration $c$ increases drawing $p$ to the $\frac{1}{G_n}$, then
the estimate (\ref{Gl1}) turns into $G_l>0$, which promises nothing as regards selectivity
gain.

\section{CONCLUSIONS AND DISCUSSION}

We have considered here a set of identical receptor proteins belonging to a single ORN
as converging on its interior and made some conclusions about sensitivity and selectivity
gain in the ORN itself, see Equations (\ref{Gn}), (\ref{finest}), (\ref{Gl1}). The main conclusion is that
due to convergence one may have either high selectivity or high sensitivity. Increasing
one of them results in decreasing the other one. While the sensitivity gain does not
depend on the stimulus intensity, the selectivity decreases with increasing concentration,
see Equations (\ref{finest}), (\ref{Gl1}). This conforms with experimental observations,
\cite{Duchamp1984,Duchamp-Viret1990}. Concentration at which a selectivity gain might be
expected, must be sub-threshold: the firing threshold can be achieved due to fluctuations,
but the time-averaged concentration is less than the threshold one. This strong limitation
on the concentration range may cast doubt on a possibility that the mechanism discussed here 
could
operate in a real biological system. In this connection I would like to say that the suitable
concentration depends on the threshold $N_0$, which itself could be adjustable. The 
evident mechanisms for this are adaptation and inhibition of the ORN. Interesting, that in the
olfactory system, ORN's output is subjected to feedback presynaptic inhibition, \cite{McGann2013}.
For the secondary neurons, this inhibition acts similarly as elevation of ORN's firing threshold.

Conclusions made here about the trade-off between sensitivity and selectivity 
and the selectivity gain itself are
obtained by means of mathematical analysis of the binding-releasing stochastic process,
\cite{Vidybida2003a}. 

As to my knowledge, no experimental attempt was made to compare
the ORN selectivity with that of its receptor proteins. This is not surprising because
measuring selectivity of a receptor protein belongs to chemistry, while discriminating
ability of ORN --- to sensory biology. 
The selectivity gain predicted in \cite{Vidybida2003a} and
 here is made possible due to the threshold manner 
of the ORN response to converging inputs from its receptors.
 This offers an experimental procedure for checking the prediction.
Namely, the selectivity of individual receptor protein ($\mu$) can be estimated while measuring
the ORN output in the form of receptor potential, with spike-triggering mechanism blocked.
The ORN selectivity ($\delta$) can be estimated through mean firing rate as defined in
(\ref{delta}).

\section{ACKNOWLEDGMENTS}
This research was supported by theme grant of department of physics and astronomy of NAS Ukraine "Dynamic formation of spatial uniform structures in many-body system" PK0118U003535.

Also, I would like to thank to the 30th Annual Biophysics Congress (International), organized by Turkish Biophysical society, and to Prof. Mehmet Can Akyolcu personally for organizing a nice Congress and
 inviting me.


%

\end{document}